\documentclass[12pt]{article}
\usepackage[dvips]{epsfig}
\usepackage{graphicx,natbib,amsmath,amsfonts,fullpage,rotating,setspace}
\usepackage{hyperref}

\begin{document}

\large
\onehalfspacing

\baselineskip 17pt
\parindent 10pt
\parskip 9pt

\begin{center}
{\Large {\bf  A stability analysis of the power-law steady state of marine size spectra}}\\ \vspace{1cm} {\large
Samik Datta*, Gustav W. Delius*, Richard Law*, Michael J. Plank**}
\\ \vspace{3mm} {\em* Departments of Biology and Mathematics,\\ University of York, York YO10 5DD, U.K.}
\\ \vspace{3mm} {\em** Department of Mathematics and Statistics, University of Canterbury,\\ Christchurch, New Zealand}
\footnote{emails: {\tt* sd550, gwd2, rl1 @york.ac.uk, ** M.Plank@math.canterbury.ac.nz} }
\end{center}

\begin{abstract}
This paper investigates the stability of the power-law steady state often observed in marine ecosystems. Three dynamical systems are considered, describing the abundance of organisms as a function of body mass and time: a ``jump-growth'' equation, a first order approximation which is the widely used McKendrick-von Foerster equation, and a second order approximation which is the McKendrick-von Foerster equation with a diffusion term. All of these yield a power-law steady state. We derive, for the first time, the eigenvalue spectrum for the linearised evolution operator, under certain constraints on the parameters. This provides new knowledge of the stability properties of the power-law steady state. It is shown analytically that the steady state of the McKendrick-von Foerster equation without the diffusion term is always unstable. Furthermore, numerical plots show that eigenvalue spectra of the McKendrick-von Foerster equation with diffusion give a good approximation to those of the jump-growth equation. The steady state is more likely to be stable with a low preferred predator : prey mass ratio, a large diet breadth and a high feeding efficiency. The effects of demographic stochasticity are also investigated and it is concluded that these are likely to be small in real systems.

\addvspace{0.2 in}

\end{abstract}

\small

\section{Introduction}\label{section.intro}

It is well established that marine ecosystems often show roughly equal abundances of biomass in logarithmically increasing weight intervals, when organisms are identified by body mass rather than by species identity \citep{sheldon:72, boudreau:92}. This is equivalent to a power-law for the abundance density as a function of body mass with exponent of approximately $-2$. Alternatively, plotting log(abundance) against log(mass) gives a ``size spectrum'' \citep{sheldon:67, platt:78} which is approximately linear with gradient near to $-1$.

\par This empirical pattern has motivated a programme of theoretical research. Silvert and Platt \citeyearpar{silvert:78, silvert:80} developed a size-dependent partial differential equation modelling growth and death in a size spectrum, and established the existence of a power-law steady state. The power-law steady state has also been shown in systems where predators are allowed to eat any prey smaller than themselves \citep{camacho:01}. When predators are assumed to be more selective (i.e. eating only a certain range of prey), the existence of a power-law steady state has also been proven, using an integro-differential equation for the model instead of a partial differential equation; the exponent generally depends on assimilation efficiency, external mortality and predator-prey interaction rates \citep{benoit:04}. In these and other studies \citep[e.g.][]{andersen:06, blanchard:09, law:09}, the McKendrick-von Foerster equation is commonly used. However, a derivation from a stochastic model of predation leads to a more general equation \citep{datta:10}, which we will refer to as the ``jump-growth'' equation in the following analysis. The McKendrick-von Foerster equation is the first order approximation (in an infinite series) to the jump-growth equation when prey are typically much smaller than predators. The second order approximation brings a diffusion term into the McKendrick-von Foerster equation \citep{datta:10}, the effects of which have not previously been studied.

\par Marine biologists need to understand the resilience of the power-law steady state to perturbations caused by fishing and natural phenomena, such as springtime plankton blooms. For instance, it has been shown that fishing increases the temporal variability in abundance of marine species \citep{hsieh:06, anderson:08}. Fundamental to this understanding are the stability properties of the power-law steady state, about which very little is known. We do know from recent numerical studies on the jump-growth equation and the McKendrick-von Foerster equation that there is a bifurcation from a stable power-law steady state to a travelling-wave attractor under certain parameter conditions \citep{law:09, datta:10}. However, the only stability analysis we are aware of assumed growth to be independent of prey density \citep{arino:04}, thereby excluding a key predator-prey interaction at the heart of the dynamics. The power-law steady state plays a pivotal role in marine ecosystems, and it is essential to understand the factors that contribute to its stability and instablity.

This paper provides the first detailed stability analysis on the jump-growth equation and its low order approximations, the McKendrick-von Foerster equation and the McKendrick-von Foerster equation with diffusion. It is also the first analysis of the effects of including the second order diffusion term in the McKendrick-von Foerster equation, and of the effects of demographic noise on the stable power-law steady state. The results show that the first order approximation is unstable, whereas the second order approximation can be stable, and gives a much better approximation to the jump-growth equation. The steady state is shown to be more likely to be stable when the preferred predator : prey mass ratio is reduced and the diet breadth and the feeding efficiency are increased.

\par For readers interested in the mathematical derivation of the perturbation equations and eigenvalue spectra, Section \ref{section.maths} shows the necessary steps taken. However, for those more interested in the results of the stability analyses, Section \ref{section.results} shows the behaviour of the three models, and reading Section \ref{subsection.models} should provide sufficient background reading to understand the different models used.
\section{Analysis of the power-law steady state}\label{section.maths}

\subsection{Three models of predation}\label{subsection.models}

The analysis focuses on perturbations around the power-law steady state of three equations: the jump-growth equation \eqref{djge}, the McKendrick-von Foerster equation \eqref{mvf} and the McKendrick-von Foerster equation with diffusion \eqref{mvfd}. These equations describe the rate of change in the density of organisms of weight $w$, which we call $\phi(w)$, with dimensions M$^{-1}$L$^{-3}$, where M is the mass dimension and L is the length dimension. This density is with respect to both mass and volume, so the number of organisms in a volume $V$ with weight between $w$ and $w+dw$ is $V\phi(w)dw$. The first equation is based on the jump-growth equation of \citet{datta:10},
\begin{eqnarray}\label{djge}
\frac{\partial\phi(w)}{\partial t} &=&\int\left(-T(w,w')\phi(w)\phi(w')-T(w',w)\phi(w')\phi(w) \right. \nonumber \\
                                   &&\ \ \ \ \left.+T(w-Kw',w')\phi(w-Kw')\phi(w')\right)\ dw'-\mu\phi(w).
\end{eqnarray}
$T(w,w')$ is proportional to the feeding rate of individuals of weight $w$ on individuals of weight $w'$, and $0 < K < 1$ is the conversion efficiency of biomass from prey to predator \citep{law:09}. There are three ways in which a feeding event can result in a change in the density of individuals at a given weight $w$, corresponding to the three terms in the integrand. The first term represents the loss of individuals of weight $w$ due to growth to a larger size (predation of $w$ upon $w'$), the second term the loss of individuals of weight $w$ due to death (predation of $w'$ upon $w$), and the third term the gain of individuals of weight $w$ due to growth from from a smaller size (predation of $w-Kw'$ on $w'$). Here we have also included a linear natural death rate $\mu$ (with the dimension of inverse time) to allow for other sources of mortality.

\par A Taylor expansion of the third term in the jump-growth equation in powers of $K$ gives an infinite series of approximations to the full jump-growth equation \citep{datta:10}. Expanding up to and including terms linear in $K$ gives our second model, the McKendrick-von Foerster equation,
\begin{eqnarray}\label{mvf}
\frac{\partial\phi(w)}{\partial t} &=& -\int T(w',w)\phi(w)\phi(w')\ dw' \\
                                   && -\frac{\partial}{\partial w}\int Kw'\,T(w,w')\phi(w)\phi(w')\ dw'-\mu\phi(w) \nonumber
\end{eqnarray}
and including terms quadratic in $K$ gives our third model,
\begin{eqnarray}\label{mvfd}
\frac{\partial\phi(w)}{\partial t} &=& -\int T(w',w)\phi(w)\phi(w')\ dw'  \\
                                   && -\frac{\partial}{\partial w}\int Kw'\,T(w,w')\phi(w)\phi(w')\ dw' \nonumber \\
                                   && +\frac{1}{2}\frac{\partial^2}{\partial w^2}\int (Kw')^2\,T(w,w')\phi(w)\phi(w')\ dw'-\mu\phi(w), \nonumber
\end{eqnarray}
which we will refer to as the McKendrick-von Foerster equation with diffusion. Note that, as in equation \eqref{djge}, a linear death rate $\mu$ has been included in these two approximations.

\par We assume a feeding kernel of the form
\begin{equation}\label{kernel1}
T(w,w') = Aw^{\alpha}s\left(\frac{w}{w'}\right)
\end{equation}
where $A$ is the predator search volume per unit mass$^{-\alpha}$ per unit time, $\alpha$ is the predator search exponent, calculated to have a value of approximately 0.8 \citep[see][]{ware:78}, and $s(w/w')$ is the feeding preference function, centred around some preferred predator : prey mass ratio $B$. To make analytical progress in this paper we assume that $\alpha = \gamma-1$, where $\gamma$ is the exponent of the power-law steady state ($\approx$ 2). This assumption then has the consequence that the steady state is a power-law (see below). In addition, the eigenvalue spectrum can then be written as a closed form expression and its properties analysed. Although probably not realistic from a biological point of view (discussed in Section \ref{section.discussion}), the assumption places stability analyses of size spectra on a firm mathematical foundation and provides a basis from which exploration of a broader class of systems can begin.

\par Section \ref{subsection.ss} defines the power-law steady state for the jump-growth equation (\ref{djge}) and its two approximations (\ref{mvf}) and (\ref{mvfd}). Section \ref{subsection.pert} develops equations for the dynamics of small perturbations to this steady state and Section \ref{subsection.eigenvalues} gives explicit equations for the eigenvalue spectra. In Section \ref{subsection.fluctuations}, the effect of demographic noise on the system at steady state is investigated. Finally, Section \ref{subsection.gaussian} incorporates a Gaussian feeding preference for predators.

\subsection{The power law steady state}\label{subsection.ss}

The steady state for equations \eqref{djge}, \eqref{mvf} and \eqref{mvfd} is given by
\begin{equation}
\hat{\phi}(w) = \phi_0w^{-\gamma}
\end{equation}
where $\phi_0$ is a constant. Below, it helps to transform the variable $w$ to a dimensionless log weight variable $x = \ln(w/w_0)$ (for some arbitrary weight $w_0$). For analysing the steady state of the jump-growth equation (\ref{djge}) it is convenient to change the integration variable of each of the three terms to the predator : prey mass ratio, which leads to the transformed equation
\begin{eqnarray}\label{djgev}
\frac{\partial v(x)}{\partial t} &=& \hat{A}\int \,s(e^r)\Bigg(-e^{\alpha r}v(x)v(x-r)-v(x)v(x+r) \nonumber \\
                      && \ \ \ \ \ \ \ \ \ \ \ \ \ \ \ \ \ +e^{\alpha(r+\psi(r))}v(x-\psi(r))v(x-r-\psi(r))\Bigg)\ dr-\mu v(x),
\end{eqnarray}
where we have used equation \eqref{kernel1} for the feeding kernel with $\alpha = \gamma-1$. Here $v(x)$ has the property that  $e^{-(\alpha+1)x}v(x)\,dx = \phi(w)\,dw$ and has dimensions L$^{-3}$, $\hat{A} = A{w_0}^\alpha$, and $r$ is the log of the predator : prey mass ratio with $\psi(r) = \ln(1+Ke^{-r})$. In the transformed jump-growth equation \eqref{djgev}, the steady state is simply given by
\begin{equation}
v(x) = v_0,
\end{equation}
where $v_0=\phi_0{w_0}^{-\alpha}$ is a constant. Substituting this into equation \eqref{djgev} we get the steady state condition,
\begin{equation}\label{ss}
\int \,s(e^r)\,\left(-e^{\alpha r}-1+e^{\alpha(r+\psi(r))}\right)\ dr-\eta = 0
\end{equation}
where $\eta = \mu/(\hat{A}v_0)$ is dimensionless. This equation implicitly determines the value of the search volume exponent $\alpha$ (and thus the steady state exponent $\gamma$) for a given choice of the parameters $K$ and $\eta$ and the feeding kernel $s(e^r)$. If we impose the conditions that predators can only feed upon prey smaller than themselves and $K\neq0$, we can prove analytically that there always exists a unique value for $\alpha$ that solves the steady state condition. Without these conditions we verify its existence and uniqueness numerically. Setting $\eta$ determines the abundance of fish at the steady state, as it contains the constant $v_0$.

\par For the McKendrick-von Foerster equation with diffusion (\ref{mvfd}), the steady state condition is
\begin{equation}\label{alphamvfd}
\int \,s(e^r)\left(-1+\alpha Ke^{(\alpha-1)r}+\alpha(\alpha-1)\frac{K^2}{2}e^{(\alpha-2)r}\right)\ dr-\eta = 0,
\end{equation}
and for the McKendrick-von Foerster equation without diffusion (\ref{mvf}), terms of order $K^2$ in equation (\ref{alphamvfd}) are ignored.

\subsection{Perturbations around the steady state of the jump-growth equation}\label{subsection.pert}

\par We now add a small perturbation to the steady state of the jump-growth equation and observe its evolution over time. If the perturbation grows over time, then the steady state is not stable, and the system will not stay at the equilibrium; if the perturbation decays, then the steady state is locally asymptotically stable. We call the perturbation $v_0\epsilon(x,t)$ and obtain its evolution equation by substituting
\begin{equation}
v(x,t) = v_0(1+\epsilon(x,t))
\end{equation}
into equation \eqref{djgev}. We now assume that we can neglect terms of order $\epsilon^2$ because $\epsilon$ is taken to be very small. For a finite-dimensional dynamical system this can be justified rigorously using the Hartman-Grobman theorem (see e.g. \cite{kirchgraber:90}). However in an infinite-dimensional system this can be more subtle (see e.g. \cite{aulbach:93}) and we proceed formally in analogy with the finite-dimensional case. We then use condition (\ref{ss}) to eliminate terms of order $\epsilon^0$, so that only terms of $\epsilon^1$ remain. This leads to the linearised perturbation equation
\begin{eqnarray}\label{pertjg}
\frac{\partial\epsilon(x)}{\partial t}&=&\hat{A}v_0\int \,s(e^r)\Bigg(-e^{\alpha r}(\epsilon(x)+\epsilon(x-r)) \\
                      &&\ \ \ \ \ \ \ \ \ \ \ \ \ \ \ \ \ \ \ \ -(\epsilon(x)+\epsilon(x+r)) \nonumber \\
                      &&\ \ \ \ \ \ \ \ \ \ \ \ \ \ \ \ \ \ \ \ +e^{\alpha(r+\psi(r))}(\epsilon(x-\psi(r))+\epsilon(x-r-\psi(r)))\Bigg)\ dr-\mu\epsilon(x). \nonumber
\end{eqnarray}
We can change integration variables appropriately so that the right hand side of equation (\ref{pertjg}) is in the form of an integral operator acting on $\epsilon$,
\begin{equation}\label{pertjg2}
\frac{\partial\epsilon(x)}{\partial t} = \hat{A}v_0\int \epsilon(m)G(x,m)\,dm
\end{equation}
where
\begin{eqnarray}\label{convol}
G(x,m) &=&-\delta(r)\left(\int s(e^z)(e^{\alpha z}+1)dz +\mu\right)-s(e^r)e^{\alpha r}-s(e^{-r}) \nonumber \\
       &&+s(e^{z_1})K^{-1}e^{(\alpha+1)(z_1+r)}+s(e^{z_2})e^{(\alpha+1)r-z_2}.
\end{eqnarray}
Here $r = x-m$, $z_1 = \ln(K/(e^r-1))$, $z_2 = \ln(e^r-K)$ and $\delta$ represents the Dirac delta function. The integral kernel $G(x,m)$ can be thought of as an infinite-dimensional version of a matrix with indices $x$ and $m$ and the task of solving equation \eqref{pertjg2} thus reduces to finding the `eigenvectors' and `eigenvalues' of this `matrix'. To define the operator rigorously in the infinite-dimensional case we must first restrict the perturbations to the space of square-integrable periodic functions with some period $L$. On this space the operator is compact and thus it is meaningful to speak of its spectrum of eigenvalues. In the end we can then take the period $L$ to infinity.

\subsection{Eigenvalue spectra}\label{subsection.eigenvalues}

We observe that the integral kernel $G(x,m)$ depends on $x-m$ only, i.e. it is a convolution kernel. Its `eigenvectors' are given by plane waves, $\epsilon_k(x)=e^{ikx}$, for any $k\in\mathbb{R}$. We refer to $k$ as the wavenumber of the plane wave $\epsilon_k(x)$ and denote the corresponding eigenvalue as $\lambda(k)$.

\par The eigenvalues are
\begin{equation}\label{lambdajg}
\lambda(k) = \int \,s(e^r)\,\left(-e^{\alpha r}-e^{ikr}+e^{\alpha r+(\alpha-ik)\psi(r)}\right)\left(1+e^{-ikr}\right)\ dr-\eta.
\end{equation}
We refer to the values taken by $\lambda(k)$ as the eigenvalue spectrum.

\par A general perturbation can then be expanded in terms of these plane waves and its time evolution is
\begin{equation}\label{pw}
\epsilon(x,t) = \int C(k)e^{ikx+\hat{A}v_0\lambda(k)t}\ dk.
\end{equation}
The expansion coefficient function $C(k)$ is an even function because $\epsilon(x,t)$ is real. Notice that if any $\lambda(k)$ has a positive real part then perturbations grow exponentially with time (the factors $\hat{A}$ and $v_0$ are positive constants and thus do not affect the coefficient of $t$), which means that the steady state is unstable.

\par To derive the eigenvalue spectrum for the McKendrick-von Foerster equation with diffusion from equation (\ref{lambdajg}), $\psi(r)$ is expanded in powers of $K$. Taking terms up to and including $K^2$ yields
\begin{equation}\label{lambdamvfd}
\lambda(k) = \int \,s(e^r)\,\left(-e^{ikr}+K(\alpha-ik)e^{(\alpha-1)r}+\frac{K^2}{2}(\alpha-ik)(\alpha-1-ik)e^{(\alpha-2)r}\right)\left(1+e^{-ikr}\right)\ dr-\eta.
\end{equation}
As in equation (\ref{alphamvfd}), neglecting $K^2$ terms gives the corresponding eigenvalue spectrum for the McKendrick-von Foerster equation.

\par It is the real part of the eigenvalue that we are interested in, as it is the sign of this that determines whether the perturbations grow or die out over time. If, for some wavenumber $k$ Re$(\lambda(k))$ is positive, then any perturbation containing a component with this wavenumber will grow over time and thus the steady state will be unstable. If Re$(\lambda(k))$ is negative for all $k$ then all perturbations die out over time, and the steady state is stable.

\subsection{Stochastic fluctuations}\label{subsection.fluctuations}

The analysis above is concerned with the deterministic jump-growth equation (\ref{djge}) and its low-order approximations (\ref{mvf}) and ({\ref{mvfd}). In fact, equation (\ref{djge}) is the mean-field equation for a stochastic model of pairwise encounters between predator and prey \citep{datta:10}. The magnitude of the fluctuations due to the demographic noise in the stochastic model is usually a factor of $\Omega^\frac{1}{2}$ smaller than the mean-field solution, where $\Omega$ is the number of individuals in the system \citep{vankampen:92}. For marine ecosystems $\Omega$ tends to be very large, so the fluctuations will be relatively small, but they can nonetheless have important effects \citep{mckane:05}, and may significantly impact the patterns observed in empirical data.

In this section we describe how the magnitude of the stochastic fluctuations, and the correlations between the fluctuations at different body sizes, can be predicted. In order to make the following statements rigorous, one would work in terms of discrete body size intervals, but we work in the continuum formally for convenience, which gives the same results.
We let $n(x,t)$ be a random variable corresponding to the density of individuals of size $w=w_0e^x$ at time $t$. The random variable is described by the stochastic process given in previous work  \citep{datta:10}. Following the method used by \citet{vankampen:92}, we separate $n(x,t)$ into a deterministic component $v(x,t)$, which satisfies the mean-field equations studied above, and a random fluctuation component $\xi(x,t)$:
\begin{equation}
n(x,t) = Ve^{-\alpha x}\left(v(x,t) + \Omega^{-\frac{1}{2}}v_0\xi(x,t)\right).
\end{equation}
Since the focus of this paper is the stability of the steady state, we restrict attention to the case where the deterministic component is at steady state; the results are therefore only relevant in cases where the steady state is stable. The stochastic fluctuations $\xi(x,t)$ can be described by a Langevin-type equation
\begin{equation}\label{langevin}
\frac{\partial}{\partial t}\xi(x,t) = \hat{A}v_0\left(\int G(x,y)\xi(y,t)dy+\rho(x,t)\right),
\end{equation}
where the kernel $G$ is given by equation (\ref{convol}) and $\rho(x,t)$ is a null-mean noise process. Details of the derivation of this equation in the general, non-equilibrium setting may be found in \citet{datta:10}. The covariance of noise at two different body sizes is described by a covariance kernel $B(x,y) = \langle \rho(x,t)\rho(y,t)\rangle$, which is given by \citep[see][]{datta:10}:
\begin{equation}
\begin{split}
B(x,y) = e^{\alpha(x+y)}\int\Bigg(& f(x,y,z)-f(x,z,y)-f(z,y,x) \\
                                  & +\delta(x-y)\int \left(f(x,z,z')+\frac{f(z,z',x)}{2}\right) dz'\Bigg)dz,
\end{split}
\end{equation}
where
\begin{equation}
f(x,y,z)=e^{-\alpha(x+y)}\left(k(x,y,z)+k(y,x,z)\right)
\end{equation}
and
\begin{equation}
k(x,y,z)=e^{\alpha x}s\left(e^{x-y}\right)\delta(z-x-\psi(x-y)).
\end{equation}
The covariance $\langle \xi(x,t) \xi(y,t)\rangle$ of the fluctuations at logarithmic body sizes $x$ and $y$ satisfies
\begin{equation}\label{eq.corr}
\frac{\partial}{\partial t}\langle \xi(x,t) \xi(y,t)\rangle = \hat{A}v_0 \left(\int\left(G(x,z)\langle \xi(z,t) \xi(y,t)\rangle
+G(y,z)\langle \xi(z,t) \xi(x,t)\rangle \right)dz+ B(x,y)\right).
\end{equation}
In the steady state, the time derivative on the left-hand side vanishes and thus the covariance function $\langle \xi(x,t) \xi(y,t)\rangle$ can be calculated by setting the right-hand side to zero which results in a linear equation to be solved. 
We present the numerical results of this in section \ref{subsection.corrplot}. In order to verify these, we also carry out stochastic simulations of the number $n_i(t)$ of individuals in log weight bracket $[x_i,x_{i+1}]$ for $-4\le x\le 4$ (outside this size range, the spectrum is assumed to remain at steady state). We approximate the number $R_{ij}$ of individuals in bracket $i$ that eat an individual in bracket $j$ during a short time $\delta t$ as a Poisson random variable \citep[see][for details]{datta:10}. The mean of $R_{ij}$ is given by
\begin{equation}
V^{-1} T(w_0 e^{x_i}, w_0 e^{x_j}) n_i(t) n_j(t) \delta t.
\end{equation}
The fluctuation $\xi(x_i,t)$ is computed from the difference between $n_i(t)$ and its equilibrium value. The covariance $\langle \xi(x_i,t) \xi(x_j,t)\rangle$ is then obtained by averaging $\xi(x_i,t) \xi(x_j,t)$ over a large number of successive time points. In the long term, this gives the same result as the ensemble average of $\xi(x_i,t) \xi(x_j,t)$ provided the stochastic process is ergodic.

% The quantity $\langle \xi(x,t) \xi(y,t)\rangle$ measures the correlation between the stochastic fluctuations at two different body sizes at the same time. Fluctuations may also be correlated across different times, but an investigation of this is beyond the scope of this paper. Interaction between the demographic stochasticity inherent in the system and its natural frequency could cause these correlations to be significant \citep{mckane:05}. However, this phenomenon was observed neither in the stochastic simulations described above, nor in previous results from a similar individual-based model \citep{law:09}, even with a relatively small system size $\Omega$.

\subsection{Gaussian feeding preference}\label{subsection.gaussian}

\par Organisms do not eat indiscriminately; here we assume that they feed at some preferred prey size (in relation to their own size), and a range of sizes around this preferred size. To reflect this, a suitable preference function is a Gaussian feeding preference, with peak at $\beta$ and width proportional to $\sigma$ \citep{andersen:06, law:09}. This can be represented by the following form for $s(e^r)$,
\begin{equation}\label{gaussian}
s(e^r) = \frac{1}{\sqrt{2\pi}\sigma}\cdot e^{\frac{-(r-\beta)^2}{2\sigma^2}}.
\end{equation}
In theory this function allows predators to eat prey larger than themselves (i.e. is non-zero for $r<0$), although for realistic sets of parameter values $s(e^r)$ is typically negligible for $r<0$.

\par The eigenvalue spectrum for the jump-growth equation (\ref{lambdajg}) with the Gaussian preference function (\ref{gaussian}) unfortunately does not have a closed form. In contrast, the eigenvalue spectra for the McKendrick-von Foerster equation without and with diffusion can be determined analytically. Defining
\begin{eqnarray}
R_n &=& (\alpha-n)\left(\beta+\frac{1}{2}\sigma^2(\alpha-n)\right) \\
I_n &=& k\left(\beta+\sigma^2(\alpha-n)\right),
\end{eqnarray}
and taking the steady state condition \eqref{alphamvfd} into account, the eigenvalue spectrum for the equation with diffusion is
\begin{eqnarray}\label{lambdamvfdre}
\text{Re}(\lambda(k)) = e^{-\frac{1}{2}\sigma^2k^2}&&\Bigg[-\cos(k\beta)+Ke^{R_1}\left(\alpha\cos(I_1)-k\sin(I_1)\right) \\
&& +\frac{K^2}{2}e^{R_2}\left((\alpha(\alpha-1)-k^2)\cos(I_2)-k(2\alpha-1)\sin(I_2)\right)\Bigg]-\frac{K^2}{2}e^{R_2}k^2. \nonumber	
\end{eqnarray}
The diffusion term is removed by excluding terms of order $K^2$ in equation \eqref{lambdamvfdre}. An important difference between the two approximations is that there must always exist values of $k$ for which Re$(\lambda(k))$ is positive in the eigenvalue function for the McKendrick-von Foerster equation. Consequentially the McKendrick-von Foerster equation will never give a stable spectrum. In contrast, the McKendrick-von Foerster equation with diffusion contains a non-oscillatory term in $k$ which is negative and increases in magnitude as $k$ increases. This has the effect of making the real parts of the eigenvalues more negative for higher values of $k$.  For both approximations the oscillatory terms are damped by a factor of $e^{-\frac{1}{2}\sigma^2k^2}$. Equation \eqref{lambdamvfdre} is analysed in greater detail in Section \ref{section.results} to explain observed patterns in the behaviour of eigenvalue spectra when altering parameters.

\par Using the steady state condition \eqref{ss}, it can be shown for the jump-growth equation that Re$(\lambda(0)) = \eta$. This result also applies to both of the approximations. Thus, for any positive $\eta$, Re$(\lambda(k))$ must be positive at $k = 0$, and as $\lambda(k)$ given in equation \eqref{lambdajg} is continuous, there exists a neighbourhood around $k=0$ where Re$(\lambda(k))>0$. Therefore there will be a range of wavenumbers $k$ for which perturbations $e^{ikx}$ will destabilise the steady state. However, we only expect our model to be realistic for a range of body weights spanning around 12 orders of magnitude \citep{cohen:03} and therefore should ignore perturbations with a wavelength longer than this, i.e. those with wavenumbers smaller than about $k\approx 0.2$.

\section{Results}\label{section.results}
\begin{figure}
	\centering
		\includegraphics[width=1\textwidth]{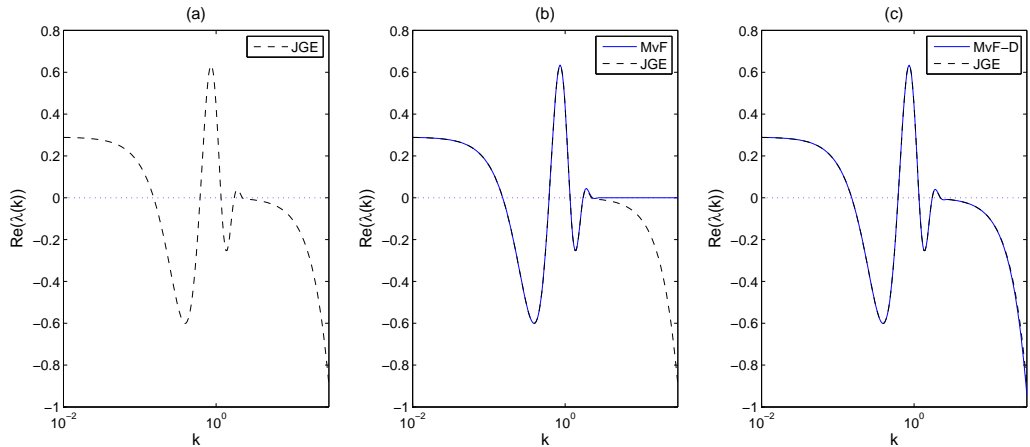}
	\caption{The eigenvalue spectra for the (a) jump-growth equation (JGE), (b) McKendrick-von Foerster equation (MvF) and (c) McKendrick-von Foerster equation with diffusion (MvF-D) when using a Gaussian feeding preference. Note that $\eta$ has been set to give a steady state exponent of roughly 2.3 for all three spectra. Parameter values $K = 0.2,\ \beta = 5,\ \sigma = 1.5,\ \eta = 0.290$, $\gamma = 2.30$.}
	\label{fig:gaussiancomp}
\end{figure}

\subsection{Eigenvalue spectra of the three models}\label{subsection.spectra}

To evaluate the eigenvalue spectra, we use the Gaussian feeding preference \eqref{gaussian}, for given values for the parameters $K$, $\beta$, $\sigma$, $\eta$. Where possible we keep these parameters biologically reasonable and close to values from previous studies \citep{andersen:06}. We use values of $K=0.2$, $\beta=5$ and $\sigma=1.5$ as a base parameter set, and investigate the effects of changing these parameters. For this base parameter set, the steady state exponent $\gamma$ is equal to 2.27 when $\eta=0$ (i.e. no external mortality) and the value of $\gamma$ increases with $\eta$. The values of $\gamma$ used in the numerical plots mostly lie in the empirical range of 2.2 to 3.25 reported by \citet{blanchard:09}. Values of the wavenumber $k$ are taken over a range from 0 to 30, as the interesting behaviour of the eigenvalue spectra is seen in this frequency range. Note that the expressions for Re$(\lambda(k))$ are even in $k$ for the three models, so the plotting of negative values of $k$ is unnecessary. We often plot the eigenvalue spectra over a logarithmic $k$-axis to make it easier to see the details at small $k$.

Examples of the eigenvalue spectra of the jump-growth equation and its two approximations (all computed numerically using the preference function \eqref{gaussian}) are compared in Figure \ref{fig:gaussiancomp}.
All three spectra are close to $\eta$ for small $k$, as expected from Section \ref{subsection.gaussian}. Both approximations are close to the jump-growth equation for low values of $k$, but as $k$ gets larger only the McKendrick-von Foerster equation with diffusion follows the jump-growth equation closely. This is expected from equation \eqref{lambdamvfdre} because the diffusion term is needed to make the eigenvalue spectrum more negative with increasing $k$. Adding the diffusion term gives a better approximation to the full jump-growth model. Thus the properties of equation \eqref{lambdamvfdre} will be used to gain insight into the behaviour of the eigenvalue spectra of the jump-growth equation in the subsequent sections.

\par The power-law steady state is unstable for all three models in this example, because all three spectra contain eigenvalues with a positive real part (the maximum occurring at $k\approx 0.861$). The tendency for unstable steady states to emerge in our analysis will be discussed in Section \ref{section.discussion}.

\par The different behaviours of the two approximations are not just limited to a Gaussian feeding preference; similar results have also been obtained when using a step function for the feeding preference. This has the form
\begin{equation}\label{step}
s(e^r) =
  \left\{
  \begin{array}{ll}
    \frac{1}{2\sigma}  & \text{if}\ \beta-\sigma \leq r \leq \beta+\sigma           \\
    0                                   & \mbox{otherwise}
  \end{array}
  \right.
\end{equation}
and is a rectangular kernel, with midpoint $\beta$, width $2\sigma$ and height 1/(2$\sigma$). It is worth noting that although behaviour similar to Figure \ref{fig:gaussiancomp} is observed, the oscillations are not damped exponentially, and oscillations are observed at all values of $k$.

\subsection{Stable and unstable steady states}\label{subsection.stable}
\begin{figure}
	\centering
		\includegraphics[width=1\textwidth]{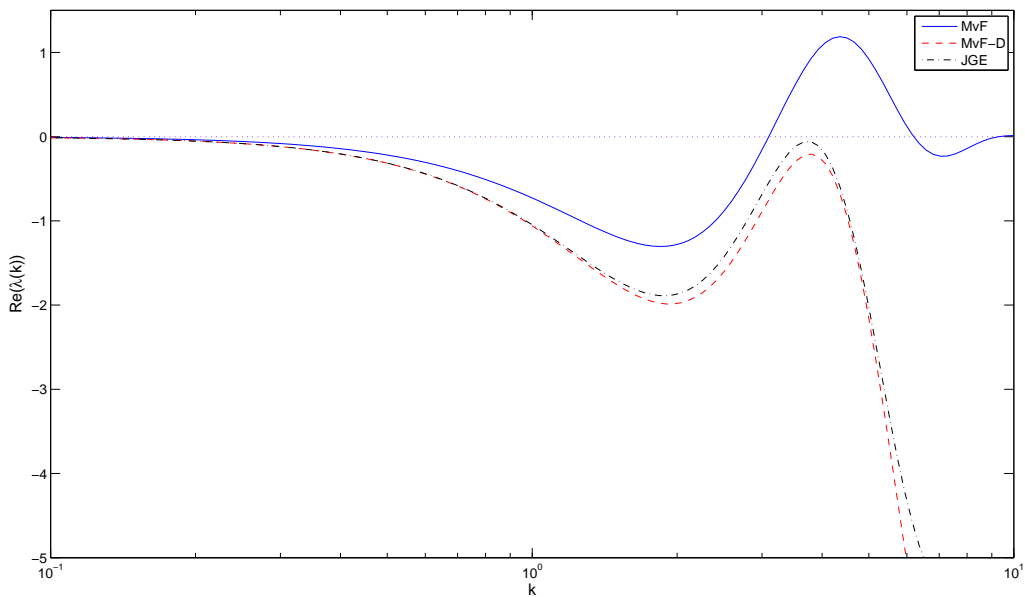}
	\caption{Eigenvalue spectra for the McKendrick-von Foerster equation and McKendrick-von Foerster equation with diffusion, compared to that of the jump-growth equation. Parameter values $K = 0.8,\ \beta = 1,\ \sigma = 0.35,\ \eta = 0$, $\gamma = 2.11$.}
	\label{fig:stable}
\end{figure}

For some sets of parameter values, the steady state is stable. Figure \ref{fig:stable} gives an example, obtained by allowing a low preferred predator : prey mass ratio $\beta$, a high efficiency $K$ and a relatively large diet breadth $\sigma$.
This example is chosen to illustrate the point that the eigenvalue spectrum for the McKendrick-von Foerster equation can be misleading; the spectrum for the McKendrick-von Foerster equation without diffusion peaks at 1.19, whereas for the equation with diffusion and the jump-growth equation Re$(\lambda(k))<0$ for all $k$. The spectrum is stabilised by the non-oscillatory term introduced by the inclusion of terms of order $K^2$. The diffusion term contributes to stability and the effect of this is great enough to make a qualitative difference to the calculated stability of the steady state. As predicted in Section \ref{subsection.gaussian}, the McKendrick-von Foerster equation gives an unstable spectrum for any choice of parameter values.

\subsection{Time evolution of perturbations}\label{subsection.evolution}
\begin{figure}
	\centering
		\includegraphics[width=1\textwidth]{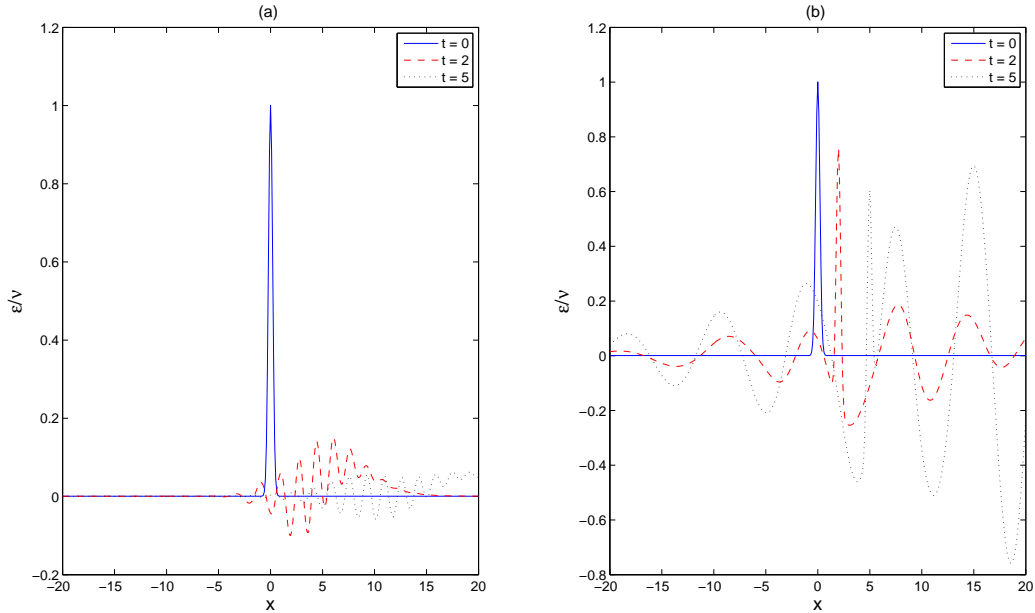}
	\caption{The time evolution of a Gaussian perturbation, $\varsigma = 0.2$, when using (a) a stable eigenvalue spectrum (parameter values $K = 0.8,\ \beta = 1,\ \sigma = 0.35,\ \eta = 0$, $\gamma = 2.11$), and (b) an unstable eigenvalue spectrum (parameter values $K = 0.2,\ \beta = 5,\ \sigma = 1.5,\ \eta = 0.290$, $\gamma = 2.30$).}
	\label{fig:packet}
\end{figure}

To show the consequences of stable and unstable steady states on the dynamics, we can examine the behaviour of a local perturbation to the size spectrum, and observe its time evolution. Assume a Gaussian perturbation with initial form
\begin{equation}
\epsilon(x,0) = \nu e^{-\frac{x^2}{2\varsigma^2}},
\end{equation}
where $\nu$ is a small constant and $\varsigma$ dictates what range of body sizes in the size spectrum are effected by the initial perturbation. This can be expanded in plane waves, rewriting $\epsilon(x,0)$ as
\begin{equation}
\epsilon(x,0) = \bar{\nu}\int_{-\infty}^\infty e^{-\frac{1}{2}\varsigma^2k^2}e^{ikx}\,dk.
\end{equation}
where $\bar{\nu} = (\nu\varsigma)/\sqrt{2\pi}$. The time dependence of this perturbation then has the following form:
\begin{equation}\label{bump}
\epsilon(x,t) = \bar{\nu}\int_{-\infty}^\infty e^{-\frac{1}{2}\varsigma^2k^2}e^{ikx+\hat{A}v_0\lambda(k)t}\,dk.
\end{equation}
We set $\varsigma$ so that the perturbation covers about one size unit on the $x$-scale, and choose units so that $\hat{A}v_0=1$. We choose to centre our perturbation around $x=0$ without loss of generality. Plotting the time evolution both for a stable spectrum (Figure \ref{fig:stable}) and an unstable spectrum (Figure \ref{fig:gaussiancomp}) using the jump-growth equation gives the two behaviours shown in Figure \ref{fig:packet}.

\par For both plots, the initial perturbation moves along the $x$-axis over time, as the organisms it contains feed on smaller organisms and grow. In the case of a stable spectrum, the perturbation gives rise to smaller peaks either side of the initial perturbation, and these all die out over time, tending to zero across the whole range of $x$. In the case of an unstable spectrum, the peaks grow over time. They develop into waves with wavenumber $\hat{k}$, where $\hat{k}$ is the most unstable node of the eigenvalue spectrum. Thus, in the case of Figure \ref{fig:packet}b, where $\hat{k} = 0.861$, the wavelength of the peaks is seen to be around $(2\pi)/\hat{k}$. Over time the peaks grow in magnitude but maintain their wavelength. The speed at which the perturbation moves through the size spectrum is determined by Im$(\lambda(\hat{k}))$.

\subsection{Changing the preferred predator : prey mass ratio}\label{subsection.beta}
\begin{figure}
	\centering
		\includegraphics[width=1\textwidth]{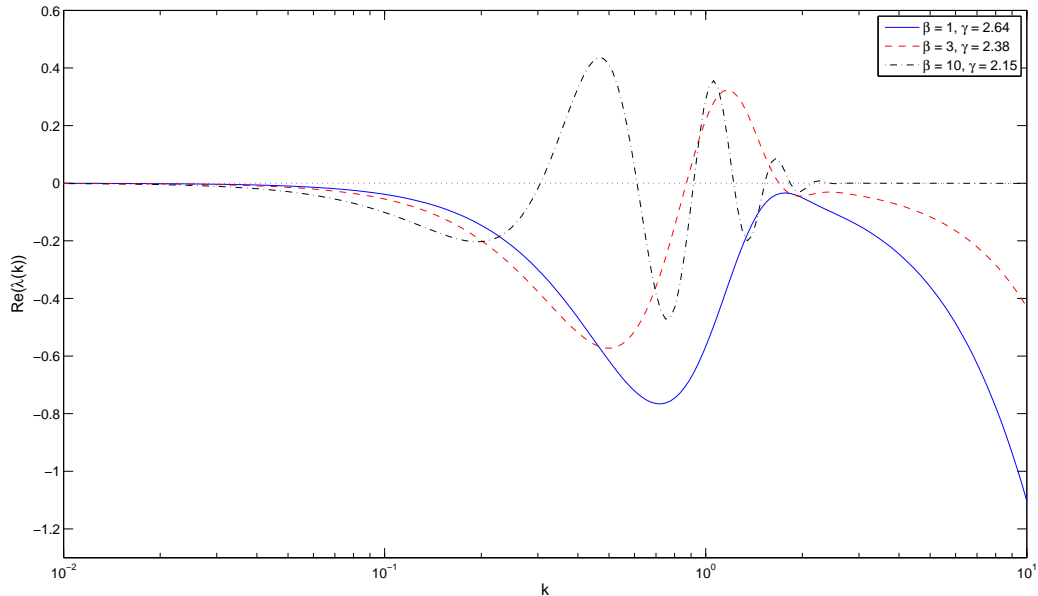}
	\caption{The eigenvalue equations for the jump-growth equation with varying logarithm of the preferred predator : prey mass ratio $\beta$. Parameter values $\sigma = 1.5, \ K = 0.2, \ \eta = 0$.}
	\label{fig:betacomp}
\end{figure}

Figure \ref{fig:betacomp} shows the effect of increasing the logarithm of the preferred predator : prey mass ratio $\beta$ on the stability of the jump-growth equation.
The maximum real part of the eigenvalues increases as $\beta$ increases, the steady state going from stability when $\beta = 1$ (i.e. Re$(\lambda(k)) < 0$ for all $k$), to instability for the larger values of $\beta$. This is in keeping with previous numerical results, where increasing $\beta$ led to a bifurcation from the power-law steady state to a travelling wave attractor \citep{law:09}, although the two results should not be directly compared because in earlier work the assumption  $\alpha = \gamma-1$ was not imposed. The changes seen in Figure \ref{fig:betacomp} as $\beta$ is increased can be understood in terms of equation \eqref{lambdamvfdre}, where $\beta$ occurs both in the $R_n$ exponential terms and in the $I_n$ cosine and sine terms. In $R_n$, $\beta$ acts to dampen the waves more as it increases, and in $I_n$, $\beta$ acts to reduce the period of the waves as it increases. Both these changes are visible in the figure. Some decrease in the exponent of the power-law steady state is also evident with increasing $\beta$ in Figure \ref{fig:betacomp}. We interpret this in biological terms as an outcome of less biomass being lost from the size spectrum as $\beta$ increases, because biomass is inefficiently consumed fewer times during its passage along the spectrum. Note that $\sigma$ has been held constant this figure, so that as $\beta$ is increased, the mean of the predator : prey feeding distribution increases but the variance remains constant.

\subsection{Changing the feeding efficiency}\label{subsection.K}
\begin{figure}
	\centering
		\includegraphics[width=1\textwidth]{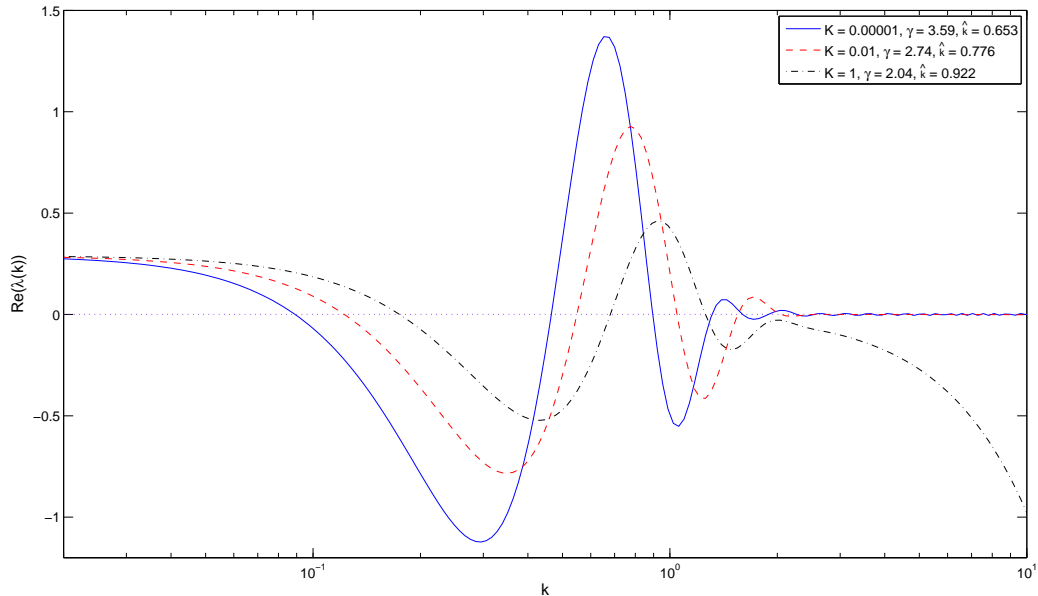}
	\caption{The eigenvalue equations for the jump-growth equation, with varying feeding efficiency $K$; $\hat{k}$ denotes the location of the most unstable node of the spectrum. Parameter values $\beta = 5, \ \sigma = 1.5, \eta = 0.290$.}
	\label{fig:kcomp}
\end{figure}

Figure \ref{fig:kcomp} shows the effect of changing the feeding efficiency $K$ on the eigenvalue spectrum.
To understand Figure \ref{fig:kcomp}, it helps to consider the limiting case of $K\rightarrow0$. Although unrealistic, because it implies no growth of organisms, the eigenvalue spectrum in equation \eqref{lambdamvfdre} is then simply a damped cosine wave: Re$(\lambda(k)) = e^{-\frac12\sigma^2k^2}\cos(k\beta)$. Consequently, the most unstable node $\hat{k}$ must be the first peak of this wave, which occurs at $\hat{k} = \pi/\beta$, equivalent to $\hat{k} = 0.628$ with the parameter values in Figure \ref{fig:kcomp}. We observe in Figure \ref{fig:kcomp} that, for small $K$ ($1\times10^{-5}$), the value of $\hat{k}$ (0.655) is close to this limiting value.

\par Corresponding to the node at $\hat{k} = \pi/\beta$, there is a dominant eigenfunction with a wavelength $2\beta$. This can be understood in biological terms as a straightforward consequence of the predator-prey interaction. A pulse perturbation from steady state that increases the density of predators at some size lowers the density of prey $e^\beta$ times smaller than themselves.  This in turn reduces the mortality rate on the prey's prey $e^{2\beta}$ times smaller than the predators, allowing their density to increase. This leads to the wavelength $2\beta$.

\par Figure \ref{fig:kcomp} also shows that, as $K$ increases, $\hat{k}$ grows and Re$(\lambda(\hat{k}))$ gets smaller. In other words, perturbations from the steady state grow more slowly and have wavelengths less than $2\beta$ as $K$ increases. In this case, a pulse increase in predator density at some body size does not remain at the same position in the size spectrum as time goes on. The predators grow as they eat, and their preferred prey body size moves along with them. This mitigates to some extent the destabilizing feedback of slow (or absent) predator growth that would continue to reduce the density of prey approximately $e^\beta$ times smaller than the predator. These results help explain the observation of \citet{law:09} that perturbations tend to have a wavelength less than $2\beta$. Notice also that the exponent of the power-law steady state becomes substantially smaller as K increases, because more biomass passes along the size spectrum to large organisms.

\subsection{Changing diet breadth}\label{subsection.sigma}

\begin{figure}
	\centering
		\includegraphics[width=1\textwidth]{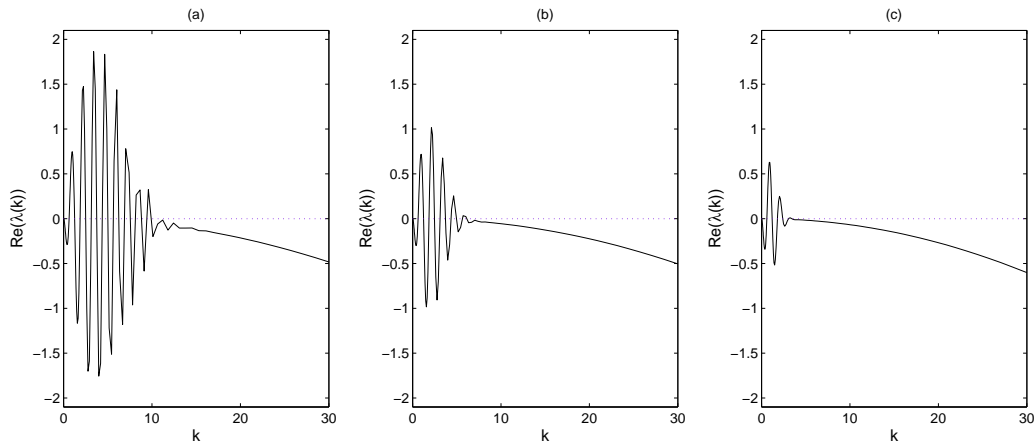}
	\caption{The eigenvalue spectra for the jump-growth equation with varying diet breadth: (a) $\sigma = 0.25$ ($k^*$ = 9.60), (b) $\sigma = 0.5$ ($k^*$ = 6.02), and (c) $\sigma = 1$ ($k^*$ = 3.41), where $k^*$ denotes the largest value of $k$ for which Re$(\lambda(k))>0$. Parameter values $K = 0.2,\ \beta = 5,\ \eta = 0,\ \gamma = 2.27$.}
	\label{fig:sigmacomp}
\end{figure}

It has been shown in earlier numerical studies that, by making the diet breadth more narrow (i.e. decreasing $\sigma$), the power-law steady state can become unstable, leading to travelling waves of abundance that move along the spectrum with time \citep{law:09, datta:10}. In the extreme case of a feeding kernel where predators only eat prey of the exact preferred mass ratio, and of no other weight (using a Dirac delta function of the form $s(e^r) = \delta(r-\beta)$ as the feeding preference), the steady state can be shown always to be unstable (proof not given here).

\par In Figure \ref{fig:sigmacomp} we investigate the effect of increasing the diet breath $\sigma$.
As $\sigma$ increases, the amplitude of oscillations at low values of $k$ decreases, and the range for which Re$(\lambda(k))$ has positive values becomes narrower; the largest value of $k$ for which Re$(\lambda(k))>0$ $k^*$ is seen to decrease as $\sigma$ increases. This is consistent with equation \eqref{lambdamvfdre}, because increasing $\sigma$ will cause the oscillations to be damped sooner by the $e^{-\frac{1}{2}\sigma^2k^2}$ term. Note that it is the change in $\sigma$ which is causing the change in the spectrum and not the steady state exponent; $\gamma$ remains at a value of approximately 2.27 in each case.

\subsection{The effects of demographic stochasticity}\label{subsection.corrplot}
\begin{figure}
	\centering
		\includegraphics[width=.65\textwidth]{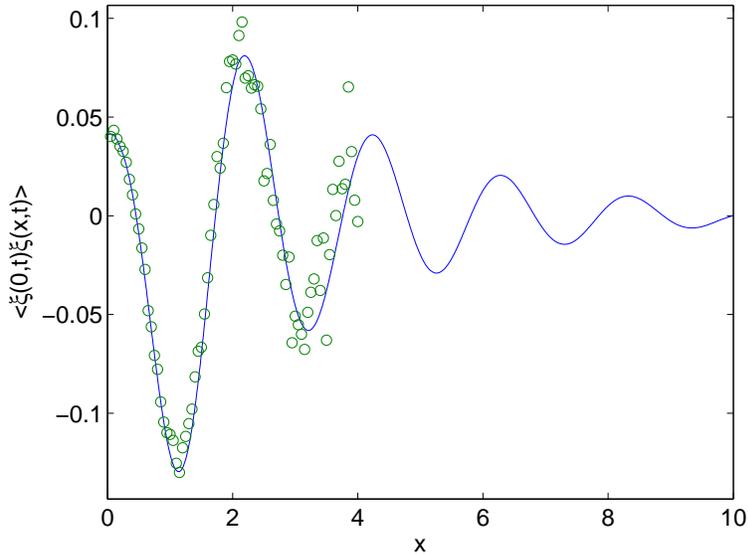}
	\caption{The covariance function $\langle \xi(0,t) \xi(x,t)\rangle$ for the fluctuations around a stable steady state due to demographic stochasticity (solid curve -- solution of equation \eqref{eq.corr}; points -- stochastic simulations). This measures the covariance between stochastic fluctuations at log body sizes $0$ and $x$; the covariance between fluctuations at log body sizes $x$ and $y$, $\langle \xi(x,t)\xi(y,t)\rangle$, is equal to $e^{\alpha x} \langle\xi(0,t) \xi(y-x,t)\rangle$. Parameter values $K = 0.8,\ \beta = 1,\ \sigma = 0.35,\ \eta = 0$, $\gamma = 2.11$.}
	\label{fig:corr}
\end{figure}

As explained in section \ref{subsection.fluctuations}, we can calculate the equal-time covariance function $\langle \xi(x,t) \xi(y,t)\rangle$ for the fluctuations in the steady state. This function describes how the fluctuations due to demographic stochasticity are correlated at different weights at steady state. It is obtained by solving the linear integral equation obtained by setting the time derivative to zero in equation \eqref{eq.corr}. To perform the calculation we used discrete weight brackets, so that the integral equation becomes a matrix equation, which we solved numerically.

Using the same parameter values as in Figure \ref{fig:stable}, the covariance function calculated from \eqref{eq.corr} (solid curve) and from stochastic simulations (points) is plotted in Figure \ref{fig:corr}. The simulations used a system size of $\Omega=10^5$ and were restricted to the size range $-4\le x\le 4$; simulating a wider size range for sufficient time was computationally prohibitive. The simulation data become increasingly noisy as $x$ increases, due to the decreasing number of individuals in a weight bracket. Nevertheless, the simulation results show good agreement with the solution of equation \eqref{eq.corr}.

The graph in Figure \ref{fig:corr} decays exponentially with distance, a typical feature of covariance functions. Superimposed on the decay is an oscillation with a wavelength of approximately $2\beta$, generated by the non-local predator-prey interaction. The reason for the oscillation is that a positive fluctuation at $x=0$ gives more food, faster growth and a negative fluctuation near $\beta$, which in turn gives less food, slower growth and a positive fluctuation near $2\beta$. %The wavelength is slightly greater than $2\beta$ because the mean distance in log-weight between a predator and its prey is slightly greater than $\beta$ (i.e. $\beta + (\gamma-1)\sigma^2$) since the prey are themselves distributed as $e^{-(\gamma-1)x}$. 

%In practice, because of the large number of individuals involved, the effects due to demographic stochasticity will be small and this was confirmed in the stochastic simulations in earlier work \citep{law:09}.

\section{Discussion}\label{section.discussion}

We have presented a local stability analysis of the power-law steady state of marine size spectra. The approach has some resemblance to the local stability analyses of steady-state food webs widely applied in ecology \citep{murray:02, rooney:06}. However, instead of having nodes representing a finite number of species, the analysis here  uses a continuous weight range corresponding to an infinite number of ``nodes'', and this gives a continuous spectrum of eigenvalues. Characterization of the eigenvalue spectrum has been carried out before \citep{arino:04}; the difference here is that we explicitly link growth of the organisms to predation, which we think is a useful step towards reality.

\par To do the analysis, the predator search exponent $\alpha$ and steady state exponent $\gamma$ have been set so that $\alpha = \gamma-1$. In addition, we assume that the rate for predation-independent death is independent of body weight. These assumptions imply that the dynamics of small perturbations are described by the convolution operator given in equation \eqref{pertjg2}, leading to a simple time dependence of the perturbations in terms of an expansion in plane waves, given in equation \eqref{pw}. In general this assumption would not be appropriate in ecological communities. The reason for using it here is that we believe it is valuable to have analytical results for this special case before beginning numerical explorations of conditions closer to those in nature.

\par The benchmark for the analysis is a jump-growth equation, obtained as the large-system limit of an underlying stochastic predation-growth process \citep{datta:10}. Importantly, the eigenvalue spectrum of the well-known, first-order approximation, the McKendrick-von Foerster equation \citep{andersen:06, law:09, blanchard:09}, exhibits a systematic departure from that of the jump-growth equation: the real parts of the eigenvalues of the former tend to zero as wavenumber increases, whereas those of the latter become increasingly negative. Therefore in our analysis the eigenvalue spectrum of the McKendrick-von Foerster equation must always contain eigenvalues with positive real parts, and must always have an unstable steady state.

\par In contrast to the first-order approximation, the eigenvalue spectrum of the second-order approximation, obtained by adding a diffusion term to the McKendrick-von Foerster equation, contains a negative term that is quadratic in the wavenumber, which makes the real parts of the eigenvalues much closer to those of the jump-growth equation.  The diffusion term is potentially important.  One consequence of it is that there can be eigenvalue spectra for which Re$(\lambda(k)) < 0$ for all wavenumbers $k > 0$, implying local stability of the steady state.  This is with the caveat that the eigenvalue spectrum tends to the natural death rate $\eta$ as the wavenumber tends to zero, so perturbations with sufficiently low wavenumbers (long wavelengths) could still destabilize the steady state.

\par The second-order approximation with diffusion has not previously been used, but would be worth considering in the future when the full jump-growth equation cannot be used. Interestingly, \citet{benoit:04} found they had to include a diffusion term in numerical integrations of the McKendrick-von Foerster equation to obtain a solution in the absence of natural mortality, although they stated that they did not understand why this should be so.  How serious the omission of the diffusion term is in practice depends on the wavenumber $k$ at which the eigenvalue spectrum peaks, because it is this wavenumber that dominates the solution in the long term. If the peak occurs at sufficiently small $k$, the effect of the negative second-order term in equation \eqref{lambdamvfdre} is small, and the standard McKendrick-von Foerster equation is reliable (Figure \ref{fig:gaussiancomp}). If the peak occurs at large  $k$, the negative second-order term in equation \eqref{lambdamvfdre} becomes significant, and inferences about stability from McKendrick-von Foerster equation may not be reliable (Figure \ref{fig:stable}). The second-order equation with diffusion itself becomes a poor approximation if the feeding preference function is set such that predators are often smaller than their prey, because the Taylor expansion of the jump-growth equation on  which it is based is no longer convergent \citep{datta:10}. However, in reality predators are almost always larger than prey, so this is not likely to be an issue.

\par Key parameters for locating the peak of the eigenvalue spectrum with respect to $k$ are the logarithm of the preferred predator : prey mass ratio $\beta$, the efficiency of mass transfer from prey to predator $K$ and the diet breadth $\sigma$. The results in Section \ref{subsection.K} suggest that predator-prey interactions would typically restrict the wavenumber $k$ at the peak to be greater than $\pi/\beta$. Overall, to get the peak of the eigenvalue spectrum at a low wavenumber where the McKendrick-von Foerster equation works best, $Ke^{-\beta}$ must be small, i.e. growth increments of predators must be small. As $\beta$ is made smaller and $K$ is made larger, the McKendrick-von Foerster approximation works less well, because it misses the stabilizing effect of the diffusion term. The diet breadth $\sigma$, also affects the shape of the eigenvalue spectrum, the main effect in equation \eqref{lambdamvfdre} being to dampen the oscillations in the real parts of the eigenvalues (Figure \ref{fig:sigmacomp}). In so doing $\sigma$ has the potential to shift positive peaks below Re$(\lambda(k)) = 0$, and hence to change an unstable steady state into a stable one. This is consistent with the results of earlier studies which have shown the stabilizing effects of broad diets \citep{law:09, datta:10}.

Random variability from one individual to another in, for example, the number and size of prey items encountered over a period of time, can have important effects in systems such as the one studied in this paper. This intrinsic demographic stochasticity is distinct from environmental stochasticity, which is not included in the current model. Usually, the relative magnitude of fluctuations due to demographic stochasticity is proportional to $\Omega^{-1/2}$, where $\Omega$ is the total number of individuals in the system \citep{vankampen:92}. However, interaction between the natural frequency of the mean-field system and intrinsic variability, which acts at all frequencies, can cause resonant amplification of demographic stochasticity \citep{mckane:05}. Stochastic effects can also cause switching between different solutions \citep[see e.g.][]{samoilov:05}, a phenomenon that cannot be investigated using the \citet{vankampen:92} approach taken in this paper, which deals with fluctuations about a mean-field solution. As seen in Figure \ref{fig:corr}, correlations between simultaneous demographic fluctuations at different body sizes do exist. These arise from the predator--prey interactions described by the feeding kernel. An investigation of correlation of fluctuations across different times is beyond the scope of this paper. However, neither resonant amplification nor switching were observed in stochastic simulations, even for relatively small system size $\Omega$, and the simultaneous correlations observed are likely to be very small for realistic system sizes.

\par A feature of the stability analysis here is that the parameter values required to achieve stability are outside the range likely to apply in marine systems (e.g. $K = 0.8$ in Figure \ref{fig:stable}). As stated above, earlier numerical integrations using the McKendrick-von Foerster equation have led to stable steady states using realistic sets of parameter values. There are, however, some important differences between the present analysis and previous work. First, real size spectra span a finite range of body sizes, about twelve orders of magnitude being realistic \citep{cohen:03}. This means that perturbations with very long wavelengths cannot occur, and corresponding to this, the wavenumber $k$ cannot be less than about 0.2. Second, the finite range calls for lower and upper bounds which are not used here. Imposing such bounds removes the exact power-law steady state, and the boundary conditions themselves influence the stability of the steady state. Third, the constraint on parameter values needed to achieve $\alpha=\gamma-1$ may exclude those values likely to lead to stability. The present study is best thought of as throwing light on the role that mortality, predation and growth play in determining stability of the power-law steady state. Other processes also leave their own footprint, and some of these increase the parameter space in which stable steady states arise \citep{capitan:10}.

\par Nonetheless, at a qualitative level, the results here are consistent with earlier observations that the steady state of marine size spectra undergoes a bifurcation from stability to instability as predator : prey mass ratio is increased and as diet breadth is decreased. The results here indicate that this is a Hopf bifurcation as a complex conjugate pair of eigenvalues cross the imaginary axis. Even without taking other major life processes into account, the analysis makes clearer what kinds of ecosystems are more vulnerable to external disturbances such as those caused by fishing and climate change. Further research should expand upon this, to better understand marine ecosystem dynamics, and better predict the potential consequences of perturbing seemingly robust ecosystems.

\addvspace{0.2 in}
\noindent \textbf{Acknowledgements}: The research was supported by a studentship to SD from the Natural Environment Research Council UK, with the Centre for Environment Fisheries and Aquaculture Science UK as the CASE partner. RL and MJP acknowledge support from the Royal Society of New Zealand Marsden fund, grant number 08-UOC-034. The research was facilitated by a Research Network Programme of the European Science Foundation on body size and ecosystem dynamics (SIZEMIC). We thank Julia Blanchard, Jennifer Burrow, Alex James, Jon Pitchford, Richard Rhodes, David Wall and the reviewers of the paper for their help and insights.

\bibliographystyle{elsart-harv}  
\bibliography{stab}

\begin{thebibliography}{25}
\expandafter\ifx\csname natexlab\endcsname\relax\def\natexlab#1{#1}\fi
\expandafter\ifx\csname url\endcsname\relax
  \def\url#1{\texttt{#1}}\fi
\expandafter\ifx\csname urlprefix\endcsname\relax\def\urlprefix{URL }\fi

\bibitem[{Andersen and Beyer(2006)}]{andersen:06}
Andersen, K.~H., Beyer, J.~E., 2006. Asymptotic size determines species
  abundance in the marine size spectrum. The American Naturalist 168, 54--61.

\bibitem[{Anderson et~al.(2008)Anderson, Hsieh, Sandin, Hewitt, Hollowed,
  Beddington, May, and Sugihara}]{anderson:08}
Anderson, C. N.~K., Hsieh, C., Sandin, S.~A., Hewitt, R., Hollowed, A.,
  Beddington, J., May, R.~M., Sugihara, G., 2008. Why fishing magnifies
  fluctuations in fish abundance. Nature 452~(7189), 835--839.

\bibitem[{Arino et~al.(2004)Arino, Shin, and Mullon}]{arino:04}
Arino, O., Shin, Y., Mullon, C., 2004. A mathematical derivation of size
  spectra in fish populations. Comptes Rendus Biologies 327~(3), 245--254.

\bibitem[{Aulbach and Garay(1993)}]{aulbach:93}
Aulbach, B., Garay, B.~M., 1993. Linearizing the expanding part of
  noninvertible mappings. {ZAMP} Zeitschrift fuer angewandte Mathematik und
  Physik 44~(3), 469--494.

\bibitem[{{Beno\^it} and Rochet(2004)}]{benoit:04}
{Beno\^it}, E., Rochet, M., 2004. A continuous model of biomass size spectra
  governed by predation and the effects of fishing on them. Journal of
  Theoretical Biology 226~(1), 9--21.

\bibitem[{Blanchard et~al.(2009)Blanchard, Jennings, Law, Castle, {McCloghrie},
  Rochet, and {Beno\^it}}]{blanchard:09}
Blanchard, J.~L., Jennings, S., Law, R., Castle, M.~D., {McCloghrie}, P.,
  Rochet, M., {Beno\^it}, E., 2009. How does abundance scale with body size in
  coupled size-structured food webs? Journal of Animal Ecology 78~(1),
  270--280.

\bibitem[{Boudreau and Dickie(1992)}]{boudreau:92}
Boudreau, P., Dickie, L., 1992. Biomass spectra of aquatic ecosystems in
  relation to fisheries yield. Canadian Journal of Fisheries and Aquatic
  Sciences 49~(8), 1528--1538.

\bibitem[{Camacho and Sol\'e�(2001)}]{camacho:01}
Camacho, J., Sol\'e�, R.~V., 2001. Scaling in ecological size spectra.
  Europhysics Letters 55~(6), 774--780.

\bibitem[{Capitan and Delius(2010)}]{capitan:10}
Capitan, J.~A., Delius, G.~W., 2010. Scale-invariant model of marine population
  dynamics. Physical Review E 81~(6), 061901.

\bibitem[{Cohen et~al.(2003)Cohen, Jonsson, and Carpenter}]{cohen:03}
Cohen, J.~E., Jonsson, T., Carpenter, S.~R., 2003. Ecological community
  description using the food web, species abundance, and body size. Proceedings
  of the National Academy of Sciences (USA) 100~(4), 1781--1786.

\bibitem[{Datta et~al.(2010)Datta, Delius, and Law}]{datta:10}
Datta, S., Delius, G.~W., Law, R., 2010. A {Jump-Growth} model for
  {Predator�Prey} dynamics: Derivation and application to marine ecosystems.
  Bulletin of Mathematical Biology 72~(6), 1361--1382. Corrected version at
  http://arxiv.org/abs/0812.4968.

\bibitem[{Hsieh et~al.(2006)Hsieh, Reiss, Hunter, Beddington, May, and
  Sugihara}]{hsieh:06}
Hsieh, C., Reiss, C.~S., Hunter, J.~R., Beddington, J.~R., May, R.~M.,
  Sugihara, G., 2006. Fishing elevates variability in the abundance of
  exploited species. Nature 443~(7113), 859--862.

\bibitem[{Kirchgraber(1990)}]{kirchgraber:90}
Kirchgraber, U., 1990. Geometry in the neighborhood of invariant manifolds of
  maps and flows and linearization. Longman Scientific \& Technical, Harlow
  Essex England.

\bibitem[{Law et~al.(2009)Law, Plank, James, and Blanchard}]{law:09}
Law, R., Plank, M.~J., James, A., Blanchard, J.~L., 2009. Size-spectra dynamics
  from stochastic predation and growth of individuals. Ecology 90~(3),
  802--811.

\bibitem[{McKane and Newman(2005)}]{mckane:05}
McKane, A.~J., Newman, T.~J., 2005. Predator-prey cycles from resonant
  amplification of demographic stochasticity. Phys. Rev. Lett. 94, 218102.

\bibitem[{Murray(2002)}]{murray:02}
Murray, J.~D., 2002. Mathematical Biology, I: an introduction, 3rd Edition.
  Springer-Verlag, Berlin.

\bibitem[{Platt and Denman(1978)}]{platt:78}
Platt, T., Denman, K., 1978. The structure of pelagic marine ecosystems.
  Journal du Conseil International pour l�Exploration de la Mer 173, 60--65.

\bibitem[{Rooney et~al.(2006)Rooney, McCann, Gellner, and Moore}]{rooney:06}
Rooney, N., McCann, K., Gellner, G., Moore, J.~C., 2006. Structural asymmetry
  and the stability of diverse food webs. Nature 442, 265--269.

\bibitem[{Samoilov et~al.(2005)Samoilov, Plyasunov, and Arkin}]{samoilov:05}
Samoilov, M., Plyasunov, S., Arkin, A.~P., 2005. Stochastic amplification and
  signaling in enzymatic futile cycles through noise-induced bistability with
  oscillations. Proc. Natl. Acad. Sci. USA 102, 2310--2315.

\bibitem[{Sheldon and Parsons(1967)}]{sheldon:67}
Sheldon, R., Parsons, T., 1967. A continuous size spectrum for particulate
  matter in the sea. Journal of the Fisheries Research Board of Canada 24,
  909--915.

\bibitem[{Sheldon et~al.(1972)Sheldon, Prakash, and {Sutcliffe
  Jr.}}]{sheldon:72}
Sheldon, R.~W., Prakash, A., {Sutcliffe Jr.}, W.~H., 1972. The size
  distribution of particles in the ocean. Limnol. Oceanog. 17, 327--340.

\bibitem[{Silvert(1980)}]{silvert:80}
Silvert, W., 1980. Dynamic energy-flow model of the particle size distribution
  in pelagic ecosystems. In: Kerfoot, W. (Ed.), Evolution and ecology of
  zooplankton communities. University Press of New England, Hanover, New
  Hampshire and London, England, pp. 754--763.

\bibitem[{Silvert and Platt(1978)}]{silvert:78}
Silvert, W., Platt, T., 1978. Energy flux in the pelagic ecosystem: a
  time-dependent equation. Limnology and Oceanography 23, 813--816.

\bibitem[{{van {K}ampen}(1992)}]{vankampen:92}
{van {K}ampen}, N.~G., 1992. Stochastic Processes in Physics and Chemistry.
  Elsevier, Amsterdam.

\bibitem[{Ware(1978)}]{ware:78}
Ware, D.~M., 1978. Bioenergetics of pelagic fish: theoretical change in
  swimming speed and ration with body size. Journal of the Fisheries Research
  Board of Canada 35, 220--228.

\end{thebibliography}
\clearpage

%\section*{Appendix}
%\input{Analytics}

\end{document}